\documentclass[prb,twocolumn,superscriptaddress,showpacs,aps]{revtex4-1}

\usepackage{color}
\usepackage{graphicx}
\usepackage{bm}
\usepackage{amsmath}
\usepackage{float}
\usepackage{gensymb}

\begin{document}

\newcommand{\Tc}{$T_{\textrm c}$}
\newcommand{\Tn}{$T_{\textrm N}$}
\newcommand{\Sr}{SrMn$_{2}$As$_{2}$}
\newcommand{\Ba}{BaMn$_{2}$As$_{2}$}
\newcommand{\BaFeCo}{Ba(Fe$_{1-x}$Co$_{x}$)$_{2}$As$_{2}$}
\newcommand{\BaFeRh}{Ba(Fe$_{0.961}$Rh$_{0.039}$)$_{2}$As$_{2}$}

\title{Collinear antiferromagnetism in trigonal {\Sr} revealed by single-crystal neutron diffraction}

\author{Pinaki Das}
\email{pdas@ameslab.gov}
\affiliation{Ames Laboratory and Department of Physics and Astronomy, Iowa State University, Ames, Iowa 50011, USA}

\author{N. S. Sangeetha}
\affiliation{Ames Laboratory and Department of Physics and Astronomy, Iowa State University, Ames, Iowa 50011, USA}

\author{Abhishek Pandey}
\altaffiliation{Present address: Department of Physics and Astronomy, Louisiana State University, Baton Rouge, Louisiana 70803, USA}
\affiliation{Ames Laboratory and Department of Physics and Astronomy, Iowa State University, Ames, Iowa 50011, USA}

\author{Zackery A. Benson}
\affiliation{Ames Laboratory and Department of Physics and Astronomy, Iowa State University, Ames, Iowa 50011, USA}

\author{T. W. Heitmann}
\affiliation{The Missouri Research Reactor, University of Missouri, Columbia, Missouri 65211, USA}

\author{D. C. Johnston}
\affiliation{Ames Laboratory and Department of Physics and Astronomy, Iowa State University, Ames, Iowa 50011, USA}

\author{A. I. Goldman}
\affiliation{Ames Laboratory and Department of Physics and Astronomy, Iowa State University, Ames, Iowa 50011, USA}

\author{A. Kreyssig}
\affiliation{Ames Laboratory and Department of Physics and Astronomy, Iowa State University, Ames, Iowa 50011, USA}

\date{\today}

\begin{abstract}
Fe pnictides and related materials have been a topic of intense research for understanding the complex interplay between magnetism and superconductivity. Here we report on the magnetic structure of {\Sr} that crystallizes in a trigonal structure ($P\bar{3}m1$) and undergoes an antiferromagnetic (AFM) transition at {\Tn} $= 118(2)$ K. The magnetic susceptibility remains nearly constant at temperatures $T \le T_{\textrm N}$ with $\textbf{\textit{H}}\parallel \textbf{\textit{c}}$ whereas it decreases significantly with $\textbf{\textit{H}}\parallel \textbf{\textit{ab}}$. This shows that the ordered Mn moments lie in the $\textbf{\textit{ab}}$-plane instead of aligning along the $\textbf{\textit{c}}$-axis as in tetragonal BaMn$_{2}$As$_{2}$. Single-crystal neutron diffraction measurements on {\Sr} demonstrate that the Mn moments are ordered in a collinear N\'{e}el AFM phase with $180^\circ$ AFM alignment between a moment and all nearest neighbor moments in the basal plane and also perpendicular to it. Moreover, quasi-two-dimensional AFM order is manifested in SrMn$_{2}$As$_{2}$ as evident from the temperature dependence of the order parameter.
\end{abstract}



\maketitle


The recent discovery of unconventional superconductivity (SC) in Fe pnictides has led to an intense research effort aimed towards understanding their fundamental properties and the underlying mechanisms that lead to strong correlations between the lattice, charge and magnetic degrees of freedom \cite{Johnston10,Canfield10,Stewart11}. Similar to the layered cuprate superconductors, the SC in FeAs-based compounds seems to arise close to an antiferromagnetic (AFM) phase, suggesting that magnetism and SC are closely intertwined in these systems \cite{Johnston97,Maple98,Lee06,Lumsden10,Dai12}. However, the parent cuprates are insulators with strongly-correlated localized magnetic moments while the parent FeAs-based superconductors are metals with itinerant moments \cite{Johnston97,Johnston10,Si08}. MnAs-based systems form a bridge between the high-{\Tc} cuprates and the FeAs-based materials, such as in tetragonal {\Ba} that orders with a G-type AFM structure and shares the same $I4/mmm$ crystal structure as many of the Fe pnictides but manifests an insulating ground state with localized moments similar to cuprates \cite{Singh09,Johnston11}. Interestingly, unlike {\Ba}, the insulator {\Sr} crystallizes in a trigonal unit cell (space group $P\bar{3}m1$) with a corrugated honeycomb structure \cite{Brechtel78,Wang11,Sangeetha16} which yields a possibly frustrated Mn spin-system \cite{Rastelli79,Mazin13,McNally15}. This kind of system attracts a lot of attention because depending upon the strength and nature of the spin interactions, different magnetic structures, from a collinear N\'{e}el AFM phase to a magnetic spiral and even to a stripe phase with alternating ferromagnetic (FM) stripes are possible \cite{Rastelli79,Mazin13}.

 \begin{figure}
  \centering
  \includegraphics{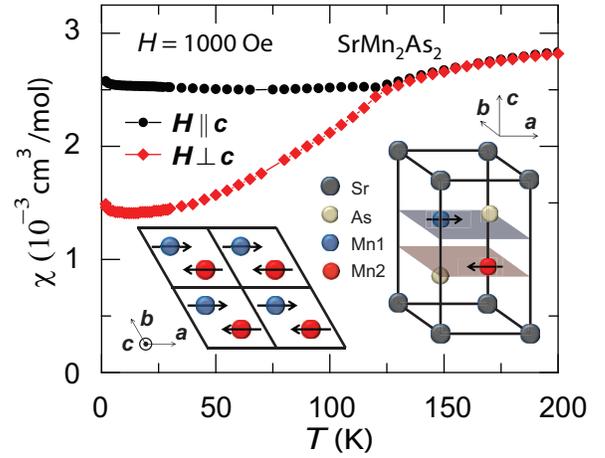}
  \caption{\label{Fig1}
           (Color online) Temperature dependence of magnetic susceptibility $\chi(T)$ for $\textbf{\textit{H}}\parallel \textbf{\textit{c}}$ and $\textbf{\textit{H}}\perp \textbf{\textit{c}}$ with $H$=1000 Oe. Two views of the magnetic structure are shown as insets. Right inset: A side view of the collinear AFM alignment of the Mn atoms at two different heights along the $\textbf{\textit{c}}$-axis marked by blue (Mn1) and red (Mn2) planes respectively, with two antiparallel spin directions. Left inset: The ordered Mn atoms as viewed along the $\textbf{\textit{c}}$-axis.}
\end{figure}

\begin{figure*}
  \centering
  \includegraphics{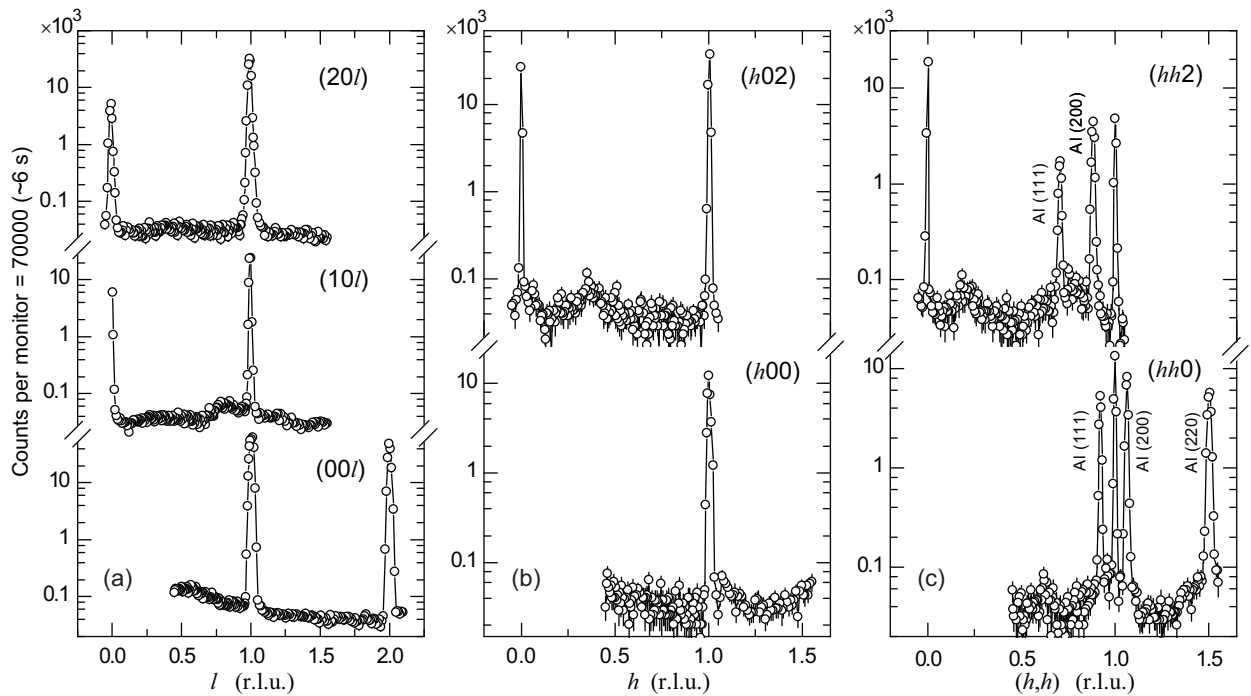}
  \caption{\label{Fig2}
           Neutron diffraction measurements at $T = 5$ K to search for magnetic Bragg peaks. (a) $l$-scans along $(h0l)$ with $h = 0$, 1, and 2. (b) $h$-scans along $(h0l)$ with $l = 0$, and 2. (c) $hh$-scans along $(hhl)$ with $l = 0$, and 2. The plots are offset along the vertical axis. Additional peaks scattered from the aluminum (Al) sample holder are observed in (c).}
\end{figure*}

Here we report single-crystal neutron diffraction studies on {\Sr} which orders in a collinear N\'{e}el AFM phase below {\Tn}~$=118(2)$~K showing that the dominant spin interactions are not frustrated. The magnetic Mn moments are antiferromagnetically aligned in the basal plane of the trigonal unit cell with an ordered moment of $3.6(2)$ $\mu_{\textrm B}$/Mn ion at $T=5$~K. The two Mn moments within the unit cell form a bilayer with antiparallel spins and can be viewed as a corrugated honeycomb lattice. High-resolution x-ray diffraction measurements were also performed to study the influence of magnetoelastic coupling in this system but no distortion of the lattice was observed down to a base temperature of 6 K within the resolution limit.


Single crystals of {\Sr} were grown out of Sn flux using conventional high-temperature solution growth techniques \cite{Canfield01,Sangeetha16}. Both magnetic susceptibility $\chi(T)$ and neutron diffraction measurements were carried out on the same single crystal of mass 44 mg, with dimensions $4.0 \times 3.0 \times 1.0$~mm$^3$. The crystals grow as flat plates with the $\textbf{\textit{c}}$-axis perpendicular to them. A Quantum Design, Inc., superconducting quantum interference device magnetic properties measurement system was used for the $\chi(T)$ measurements. Energy-dispersive x-ray measurements on the same single crystal confirmed the composition to be a pure 122 phase. Single-crystal neutron diffraction measurements were performed at the thermal triple-axis spectrometer, TRIAX, at the University of Missouri Research Reactor. Measurements were carried out with an incident energy of 14.7 meV, using S\"{o}ller collimations of $60^\prime$-$40^\prime$-sample-$40^\prime$-$80^\prime$. Pyrolytic graphite filters were placed both before and after the sample to reduce higher-order wavelengths. The sample was mounted on the cold finger of a closed-cycle helium cryostat to reach temperatures of $5$ K $\le T \le 300$ K. The lattice parameters were measured to be $a = b = 4.29(1)~\text{\AA}$ and $c = 7.24(1)~\text{\AA}$ at 5~K. Rocking scans performed through the Bragg peaks showed a full width at half maximum of $\approx$~$0.3\degree$ confirming the good mosaicity of the sample.

High-energy x-ray diffraction measurements were performed on a 1 mg single crystal at the 6-ID-D station at the Advanced Photon Source using an x-ray wavelength of $\lambda = 0.123712 ~\text{\AA}$ and a beam size of $100 \times 100 ~\mu \text{m}^2$. The sample was cooled down using a closed-cycle He cryostat. Two Be domes were placed over the sample and evacuated, and a small amount of He exchange gas was subsequently added to the inner dome for thermal equilibrium. An aluminized-Kapton heat shield also surrounded the sample and inner Be dome. The cryostat was mounted on a 6-circle diffractometer and a MAR345 image plate was used to measure the diffracted x-rays \cite{Kreyssig07}.


The $T$-dependence of the magnetic susceptibilities $\chi(T)$  with applied magnetic field ($\textbf{\textit{H}}$) along the $\textbf{\textit{c}}$-axis ($\chi_c$, $\textbf{\textit{H}}\parallel \textbf{\textit{c}}$) and perpendicular to it ($\chi_{ab}$, $\textbf{\textit{H}}\perp \textbf{\textit{c}}$) are shown in Fig.~\ref{Fig1}. For $\textbf{\textit{H}}\perp \textbf{\textit{c}}$, $\chi_{ab}$ starts to decrease rapidly below {\Tn} $=118(2)$ K with a distinct change in the slope while $\chi_{c}$ remains almost constant suggesting an antiferromagnetic (AFM) transition with the ordered moments aligned within the $\textbf{\textit{ab}}$ plane.

Theoretical calculations predict that frustration in a corrugated honeycomb lattice can result in a spiral or an incommensurate magnetic ground state \cite{Rastelli79,Mazin13}. Therefore, neutron diffraction measurements were performed to determine the magnetic structure. Experiments were carried out in two configurations -- first in the $(h0l)$ horizontal scattering plane followed by measurements in the $(hhl)$ scattering plane. Extensive measurements along the three principal directions $[00l]$, $[h00]$, and $[hh0]$ found no additional magnetic scattering beyond the magnetic contributions at the Bragg peaks corresponding to the chemical crystal structure as shown in Fig.~\ref{Fig2}. Additional diffraction peaks observed in Fig.~\ref{Fig2}(c) correspond to the scattering from the aluminum (Al) sample holder which was confirmed by rocking scans through the Al peak positions with no significant observed variation in peak intensities.


\begin{figure}
   \centering
  \includegraphics{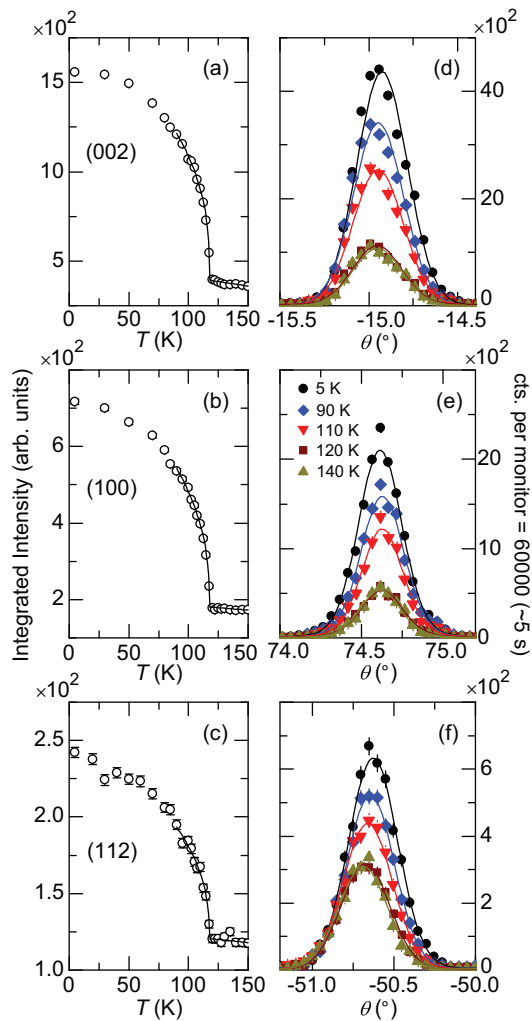}
  \caption{\label{Fig3}
           (Color online) Integrated intensities of the (002), (100) and (112) Bragg peaks as a function of temperature $T$ are shown in (a)--(c) respectively. The solid line is a power law fit given by, $I_{\textrm M}=I_0(1-T/T_{\textrm N})^{2\beta}$, for $T\ge 90$ K. (d)--(f) Rocking scans at specific temperatures [as mentioned in panel (e)] around (002), (100) and (112) peaks, respectively. The data at 120 K and 140 K overlap with each other.}
\end{figure}

As shown in Fig.~\ref{Fig3}, magnetic intensity develops only at the nuclear Bragg peak positions below {\Tn} suggesting that the magnetic unit cell is same as that of the chemical unit cell. Therefore, the magnetic structure is built by the two magnetic Mn atoms within the same unit cell only. Symmetry analysis using the program SARAh-Representational Analysis \cite{Sarah} provided distinct magnetic structures for this system which includes FM or AFM alignments with the moments aligned either along the $\textbf{\textit{c}}$-axis or in the $\textbf{\textit{ab}}$-plane. No canting of the moments is allowed by symmetry for a second-order magnetic phase transition. Since we observe magnetic intensity in $(00l)$-type Bragg peaks in our neutron diffraction measurements, we can confirm that the spins are not aligned along the $\textbf{\textit{c}}$-axis. This is due to the fact that the neutron scattering cross-section is sensitive only to the component of the ordered magnetic moment perpendicular to the scattering vector. Moreover, no indication of a FM signal was found in magnetization measurements \cite{Sangeetha16}. The only remaining possibility is a collinear N\'{e}el AFM phase with the Mn moments in the $\textbf{\textit{ab}}$-plane with $180^\circ$ AFM alignments between a moment and all nearest neighbor moments as shown in the insets to Fig.~\ref{Fig1}.

All accessible peaks in the $(h0l)$ and $(hhl)$ scattering planes have been measured and analyzed with a total of 14 independent peaks, excluding peaks with high non-magnetic intensities. Figures~\ref{Fig3}(a)--(c) show the temperature dependence of the integrated intensities of the (002), (100) and (112) peaks. Rocking scans at specific temperatures [as mentioned in panel (e)] around the (002), (100) and (112) peaks are shown in Figs.~\ref{Fig3}(d)--(f), respectively. There are no changes in the peak intensities between 120~K and 140~K but as the temperature is lowered below {\Tn}=118(2) K, the magnetic contribution sets in and the peak intensities increase.

The magnetic moment can be calculated from the integrated intensities measured by rocking scans on a series of peaks in both the $(h0l)$ and $(hhl)$ configurations. The magnetic intensity is given by \cite{Shirane}
\begin{equation}
\label{eq1}
  I_\textrm{M} = N_\textrm{M} \frac{(2\pi)^3}{v_\textrm{M}}\sum_{\textbf{\textit{G}}_\textrm{M}} \delta(\textbf{\textit{Q}} - \textbf{\textit{G}}_\textrm{M})\left | \textbf{\textit{F}}_\textrm{M} (\textbf{\textit{G}}_\textrm{M})\right |^2,
\end{equation}
where $v_\textrm{M}$ is the magnetic unit cell volume, $N_\textrm{M}$ is the number of such cells in the sample, $\textbf{\textit{Q}}$ is the scattering vector and $\textbf{\textit{G}}_\textrm{M}$ is the magnetic reciprocal-lattice vector. $\textbf{\textit{F}}_\textrm{M}$ is the magnetic structure factor and is given by
\begin{equation}
\label{eq2}
\textbf{\textit{F}}_\textrm{M} (\textbf{\textit{G}}_\textrm{M}) =  \frac{\gamma r_0}{2}\sum_{j}gf_j(\textbf{\textit{G}}_\textrm{M})\textbf{\textit{S}}_{\perp j}e^{i\textbf{\textit{G}}_\textrm{M}\cdot \textbf{\textit{d}}_j} e^{-\textbf{\textit{W}}_j},
\end{equation}
where  $\gamma = 1.193$ is the magnetic dipole moment of the neutron in units of nuclear Bohr magneton, $r_0 = 2.818 \times 10^{-15}$ m and $g$ is the spectroscopic splitting factor. $f(\textbf{\textit{G}}_\textrm{M})$ and $e^{-\textbf{\textit{W}}}$ are the magnetic form factor and the Debye-Waller factor for the magnetic Mn ions, respectively. The index $j$ is a sum over all $j^{\textrm{th}}$ Mn ions in the unit cell. The quantity $\textbf{\textit{S}}_{\perp}$ is the magnetic interaction vector and is given by $\left |  \textbf{\textit{S}}_{\perp} \right |^2 = \sum_{\alpha\beta}(\delta_{\alpha \beta} - \hat{Q}_\alpha \hat{Q}_\beta) S^{\ast}_{\alpha}S_{\beta}$. $\hat{Q}$ is a unit vector parallel to the scattering vector $\textbf{\textit{Q}}$ and the indices $\alpha$ and $\beta$ represent the $x$, $y$ and $z$ components in the summation. This reflects the fact that only the component of $\textbf{\textit{S}}$ perpendicular to $\textbf{\textit{Q}}$ contributes to the magnetic scattering amplitude. The magnetic intensities were obtained from the difference in integrated intensities between 5 K and 140 K ($>$ {\Tn}) data sets. The difference is normalized with the corresponding integrated intensities at 140 K which are purely of nuclear origin. The temperature dependence of the nuclear intensity is almost constant above {\Tn} as evident from Figs.~\ref{Fig3}(a)--(c). Thus any excess contribution of nuclear intensity at 5 K due to the Debye-Waller factor can be neglected.

Though the ordered collinear Mn moments lie in the $\textbf{\textit{ab}}$ plane of the trigonal lattice, it is not possible to uniquely determine their orientation in this plane due to the symmetry of the crystal structure. Moreover, one has to take into account the possible domain orientations in calculating the ordered moment. Considering equal domain populations and the magnetic moments aligned along the high symmetry directions, there are six possible spin directions (or domains) for the Mn moment: $\textbf{\textit{S}} = \pm [100], \pm[010]$ and $\pm[1\bar{1}0]$. The second Mn moment within the magnetic unit cell is aligned in the corresponding antiparallel direction. If $\eta_i$ is the angle between $\textbf{\textit{S}}_i$ and $\textbf{\textit{Q}}$ for the $i^{\textrm{th}}$ domain, then the average magnetic interaction vector contributing to the magnetic intensity for that particular peak is given by $\left \langle  \left |  \textbf{\textit{S}}_{\perp} \right |^2 \right \rangle = S^2(1-\left \langle \textrm{cos}^2\eta_i \right \rangle)$. The ordered moment is obtained by $\mu = gS\mu_{\textrm B}$ and was determined to be $3.6(2)$ $\mu_{\textrm B}$/Mn at 5 K. The quoted error includes the uncertainty in the domain populations. The ordered moment suggests local moment AFM behavior as observed in {BaMn$_{2}$As$_{2}$} (Ref.\citenum{Singh09}). It is somewhat lower than the nominal $5.0$ $\mu_{\textrm B}$/Mn expected for the high-spin state of Mn$^{2+}$ but is comparable to other Mn-122 compounds like {BaMn$_{2}$As$_{2}$} (Ref.\citenum{Singh09}), {BaMn$_{2}$P$_{2}$} (Ref.\citenum{Brock94}), {SrMn$_{2}$P$_{2}$} (Ref.\citenum{Brock94}), {CaMn$_{2}$Bi$_{2}$} (Ref.\citenum{Gibson15}), and {CaMn$_{2}$Sb$_{2}$} (Refs.\citenum{McNally15,Ratcliff09,Bridges09}). The reduced moment can be attributed to the strong spin-dependent hybridization between the Mn $3d$ and the As $4p$ orbitals as shown by density functional calculations \cite{An09} and to the expected quasi-two-dimensionality of the Mn--Mn spin interactions (see below). This is, however, different from the Fe-122 compounds where the Fe moment is greatly reduced due to the itinerant nature of the magnetism \cite{Johnston10,Stewart11,Si08}.

Based on the proposed domain configuration, we now calculate the expected ratio between $\chi_{ab}$ and $\chi_c$. If $\theta$ is the angle between the spin direction and the applied field, the susceptibility is given by $\chi_\theta = \chi_\parallel \mathrm{cos}^2\theta + \chi_\perp \mathrm{sin}^2\theta$, where $\chi_\parallel$ and $\chi_\perp$ are the susceptibilities parallel and perpendicular to the applied field \cite{Buschow}. For $\textbf{\textit{H}}\parallel \textbf{\textit{c}}$, $\theta = 90\degree$ for all the domains and thus $\chi_c = \chi_\perp$. To simplify the calculation for $\chi_{ab}$, we further assume $\textbf{\textit{H}}\parallel [100]$ for the in-plane susceptibility measurement. Then equal domain populations with $\theta = 0\degree$, $60\degree$ and $120\degree$ result in $\chi_{ab}=\chi_\perp/2$ at $T=0$ ($\chi_\parallel=0$ at $T=0$ for a collinear AFM). This result holds for any direction of applied field in the plane. Therefore, $\chi_{ab}/\chi_c=0.5$ at $T=0$ which is close to the experimental value of $0.58(1)$ determined from $\chi_{ab}$ and $\chi_c$ at $T = 2$~K in Fig.~\ref{Fig1}. The minor difference is probably due to a deviation from our assumption of equal domain populations.

In the following, we analyze the detailed temperature dependence of the magnetic contribution ($I_{\textrm M}$) to the Bragg peaks which is proportional to the square of the ordered moment. For $T$ close to $T_{\textrm N}$, it is predicted to have a power law behavior with a critical exponent $2\beta$ given by $I_{\textrm M}=I_0(1-T/T_{\textrm N})^{2\beta}$ (Ref. \citenum{Blundell}). Figures~\ref{Fig3}(a)--(c) show the power-law fit of the measured magnetic intensity for $T\ge 90$ K. The critical exponent $\beta$ was determined for six independent peaks as shown for the three peaks in Figs.~\ref{Fig3}(a)--(c) and was consistently found to be $0.21(3)$ which lies between the values expected for a three-dimensional (3D) Heisenberg spin system ($\beta_\textrm{3D, Heissenberg} = 0.36$) and that for a purely 2D Ising/XY system which predict critical exponents of $\beta_\textrm{2D, Ising}=0.125$ and $\beta_\textrm{2D, XY}=0.13$, respectively \cite{Blundell}. Thus {\Sr} appears to behave like a quasi-2D AFM system which is also consistent with high-$T$ magnetic susceptibility measurements [$\chi (T \ge T_{\textrm N})$]\cite{Sangeetha16}.

This quasi-2D can be explained qualitatively from the magnetic and lattice structure. Within the unit cell, the two antiparallel Mn spins lie in different planes perpendicular to $\textbf{\textit{c}}$, forming a corrugated honeycomb layer (Fig.~\ref{Fig1} insets) with a nearest-neighbor distance of $3.06(1) ~\text{\AA}$. The distance between the nearest Mn atoms between the two layers is $6.05(1) ~\text{\AA}$ which is twice the nearest Mn-Mn distance within the layer. This difference is expected to result in a weaker interplanar exchange interaction compared to intraplanar interactions leading to a quasi-2D spin system. It has been rigorously shown by Mermin and Wagner \cite{Mermin66} that a 2D arrangement of spins cannot form a long-range magnetically-ordered state except at $T=0$. However, based on recent theoretical models including the 2D quantum Heisenberg model \cite{Chakravarty89} and a combination of numerical and renormalization group arguments \cite{Taroni08}, one obtains the window for the critical exponent $\beta$ for 2D/quasi-2D systems as $\sim0.1 \le \beta \le 0.25$. This has been successful in explaining a range of systems including {La$_2$CoO$_{4}$} with a {\Tn} = 274.7(6) K and a critical exponent $\beta = 0.20(2)$ showing a crossover regime from 3D to 2D behavior \cite{Yamada89} similar to the system studied here.

Additionally, we have performed high-energy x-ray diffraction measurements to study the strength of the magnetoelastic coupling in {\Sr} which would result in the distortion of the lattice as the system enters the ordered magnetic state and breaks the trigonal symmetry. Magnetically-induced lattice distortions have been observed in other pnictide systems like in {\BaFeCo} (Ref. \citenum{Nandi10}), {\BaFeRh} (Ref. \citenum{Kreyssig10}) and many others. However, no lattice distortion was observed in {\Sr} down to 6 K and an upper limit on any possible relative lattice distortion is $3 \times 10^{-4}$ as calculated from the experimental resolution.



In conclusion, we have shown that {\Sr} orders in a collinear N\'{e}el AFM phase with the Mn spins aligned along the basal plane of the trigonal unit cell having a net ordered moment of $3.6(2)$ $\mu_{\textrm B}$/Mn which is smaller than the full high-spin value of $5.0$ $\mu_{\textrm B}$/Mn. The magnetic interaction is quasi-two-dimensional with a strong in-plane exchange interaction within the corrugated honeycomb Mn layer compared to weak interplanar interaction between the layers. Lastly, the magnetoelastic coupling does not cause a measurable lattice distortion in the ordered state.


Work at the Ames Laboratory was supported by the Department of Energy, Basic Energy Sciences, Division of Materials Sciences \& Engineering, under Contract No. DE-AC02-07CH11358. This research used resources at the University of Missouri Research Reactor. We are grateful to D. S. Robinson for support during the x-ray experiments. This research used resources of the Advanced Photon Source, a US Department of Energy (DOE) Office of Science User Facility operated for the DOE Office of Science by Argonne National Laboratory under Contract
No. DE-AC02-06CH11357.




\begin{thebibliography}{99}

\bibitem{Johnston10}
 D. C. Johnston,
 Adv. Phys. {\bf 59}, 803 (2010).

\bibitem{Canfield10}
 P. C. Canfield and S. L. Bud'ko,
 Annu. Rev. Condens. Matter Phys. {\bf 1}, 27 (2010).

\bibitem{Stewart11}
 G. R. Stewart,
 Rev. Mod. Phys. {\bf 83}, 1589 (2011).

\bibitem{Johnston97}
 D. C. Johnston,
 in {\textit{Handbook of Magnetic Materials}}, edited by K. H. J. Buschow (Elsevier, Amsterdam, 1997), Vol. 10, pp. 1--237, Chap. 1.

\bibitem{Maple98}
 M. B. Maple,
 J. Magn. Magn. Mater. {\bf 171-181}, 18 (1998).

\bibitem{Lee06}
 P. A. Lee, N. Nagaosa, and X.-G. Wen,
 Rev. Mod. Phys. {\bf 78}, 17 (2006).

\bibitem{Lumsden10}
 M. D. Lumsden and A. D. Christianson,
 J. Phys.: Condens. Matter {\bf 22}, 203203 (2010).

\bibitem{Dai12}
 P. Dai, J. Hu, and E. Dagotto,
 Nat. Phys. {\bf 8}, 709 (2012).

\bibitem{Si08}
 Q. Si and E. Abrahams,
 Phys. Rev. Lett. {\bf 101}, 076401 (2008).

\bibitem{Singh09}
  Y. Singh, M. A. Green, Q. Huang, A. Kreyssig, R. J. McQueeney, D. C. Johnston, and A. I. Goldman,
  Phys. Rev. B {\bf 80}, 100403(R) (2009).

\bibitem{Johnston11}
  D. C. Johnston, R. J. McQueeney, B. Lake, A. Honecker, M. E. Zhitomirsky, R. Nath, Y. Furukawa, V. P. Antropov, and Y. Singh,
  Phys. Rev. B {\bf 84}, 094445 (2011).

\bibitem{Brechtel78}
  E. Brechtel, G. Cordier, and H. Sch\"{a}fer,
  Z. Naturforsch. {\bf 33b}, 820 (1978).

\bibitem{Wang11}
  Z. W. Wang, H. X. Yang, H. F. Tian, H. L. Shi, J. B. Lu, Y. B. Qin, Z. Wang, and J. Q. Li,
  J. Phys. Chem. Solids {\bf 72}, 457 (2011).

\bibitem{Sangeetha16}
  N. S. Sangeetha, A. Pandey, Z. A. Benson, and D. C. Johnston (unpublished).

\bibitem{Rastelli79}
  E. Rastelli, A. Tassi, and L. Reatto,
  Physica B {\bf 97}, 1 (1979).

\bibitem{Mazin13}
  I. I. Mazin,
  (unpublished) arXiv:1309.3744.

\bibitem{McNally15}
 D. E. McNally, J. W. Simonson, J. J. Kistner-Morris, G. J. Smith, J. E. Hassinger, L. DeBeer-Schmidt, A. I. Kolesnikov, I. A. Zaliznyak, and M. C. Aronson,
 Phys. Rev. B {\bf 91}, 180407(R) (2015).

\bibitem{Canfield01}
 P. C. Canfield and I. R. Fisher,
 J. Crystal Growth {\bf 225}, 155 (2001).

\bibitem{Kreyssig07}
  A. Kreyssig, S. Chang, Y. Janssen, J. W. Kim, S. Nandi, J. Q. Yan, L. Tan, R. J. McQueeney, P. C. Canfield, and A. I. Goldman,
  Phys. Rev. B {\bf 76}, 054421 (2007).

\bibitem{Sarah}
  A. S. Wills, Physica B {\bf 276}, 680 (2000).

\bibitem{Shirane}
  G. Shirane, S. M. Shapiro, and J. M. Tranquada,
  {\textit{Neutron Scattering with a Triple-Axis Spectrometer}},
  (Cambridge University Press, Cambridge, 2002).

\bibitem{Brock94}
 S. L. Brock, J. E. Greedan, and S. M. Kauzlarich,
 J. Solid State Chem. {\bf 113}, 303 (1994).

\bibitem{Gibson15}
 Q. D. Gibson, H. Wu, T. Liang, M. N. Ali, N. P. Ong, Q. Huang, and R. J. Cava,
 Phys. Rev. B {\bf 91}, 085128 (2015).

\bibitem{Ratcliff09}
 W. Ratcliff II, A. L. Lima, A. M. Gomes, J. L. Gonzalez, Q. Huang, and J. Singleton,
 J. Magn. Magn. Mater. {\bf 321}, 2612 (2009).

\bibitem{Bridges09}
 C. A. Bridges, V. V. Krishnamurthy, S. Poulton, M. P. Paranthaman, B. C. Sales, C. Myers, and S. Bobev,
 J. Magn. Magn. Mater. {\bf 321}, 3653 (2009).

\bibitem{An09}
 J. An, A. S. Sefat, D. J. Singh, and M.-H. Du,
 Phys. Rev. B {\bf 79}, 075120 (2009).

\bibitem{Buschow}
 K. H. J. Buschow and F. R. de Boer,
 {\textit{Physics of Magnetism and Magnetic Materials}},
 (Kluwer Academic Publishers, New York, 2004).

\bibitem{Blundell}
 S. Blundell,
 {\textit{Magnetism in Condensed Matter}},
 (Oxford University Press, Oxford, 2001).

\bibitem{Mermin66}
 N. D. Mermin and H. Wagner,
 Phys. Rev. Lett. {\bf 17}, 1133 (1966).

\bibitem{Chakravarty89}
 S. Chakravarty, B. I. Halperin, and D. R. Nelson,
 Phys. Rev. B {\bf 39}, 2344 (1989).

\bibitem{Taroni08}
 A. Taroni, S. T. Bramwell, and P. C. W. Holdsworth,
 J. Phys.: Condens. Matter {\bf 20}, 275233 (2008).

\bibitem{Yamada89}
 K. Yamada, M. Matsuda, B. Keimer, R. J. Birgeneau, S. Onodera, J. Mizusaki, T. Matsuura, and G. Shirane,
 Phys. Rev. B {\bf 39}, 2336 (1989).

\bibitem{Nandi10}
 S. Nandi, M. G. Kim, A. Kreyssig, R. M. Fernandes, D. K. Pratt, A. Thaler, N. Ni, S. L. Bud'ko, P. C. Canfield, J. Schmalian, R. J. McQueeney, and A. I. Goldman,
 Phys. Rev. Lett. {\bf 104}, 057006 (2010).

\bibitem{Kreyssig10}
 A. Kreyssig, M. G. Kim, S. Nandi, D. K. Pratt, W. Tian, J. L. Zarestky, N. Ni, A. Thaler, S. L. Bud'ko, P. C. Canfield, R. J. McQueeney, and A. I. Goldman,
 Phys. Rev. B {\bf 81}, 134512 (2010).



\end{thebibliography}
\end{document}